\documentclass[fleqn,twoside]{article}
\usepackage{espcrc2}

\usepackage{graphicx}
\usepackage[figuresright]{rotating}

\newcommand{\ket}{\,\rangle}
\newcommand{\bra}{\langle \,}

\def\p{\pi}
\def\t{\tau}
\def\m{\mu}
\def\n{\nu}

\newcommand{\AmS}{{\protect\the\textfont2
  A\kern-.1667em\lower.5ex\hbox{M}\kern-.125emS}}

\title{Hadronic matrix elements for TAUOLA: $3\pi$ and $KK\pi$ channels}

\author{P. Roig\address{Instituto de F\'isica Corpuscular, IFIC, CSIC-Universitat de Val\`encia.\\ Apt. de Correus 22085, E-46071 Val\`encia, Spain}%
        \thanks{I would like to thank and congratulate Anton Poluektov and Simon Eidelman for the excellent organization of the 10th International Workshop on Tau Lepton Physics. I am indebted to J.~Portol\'es for his useful suggestions about the draft. This work has been supported in part by the EU MRTN-CT-2006-035482 (FLAVIAnet), by MEC (Spain) under grant FPA2007-60323 and by the Spanish Consolider-Ingenio 2010 Program CPAN (CSD2007-00042).}}
\begin{document}

\begin{abstract}
We emphasize that the motivation for including our hadronic matrix elements in TAUOLA is not only theoretical. We also show that our expressions describe better the $\t\to3\pi$ ALEPH data and are able to fit BABAR data on the isovector component of $e^+e^-\to KK\pi$. The theoretical foundations of our framework are the large-$N_C$ limit of QCD, the chiral structure exhibited at low energies and the proper asymptotic behaviour, ruled by QCD, that is demanded to the associated form factors.
\vspace{1pc}
\end{abstract}

\maketitle

\section{Introduction}
Among the many interesting applications that Tau Physics has \cite{Pich:2007cu}, we deal with the hadronization problem of QCD currents. Being half of the decay a semileptonic clean and controlled process, we can focus on the non-perturbative dynamics of the strong interaction.\\
\indent The decay amplitude for the considered decays may be written as:
\begin{equation} \label{Mgraltau}
\mathcal{M}\,=\,-\frac{G_F}{\sqrt{2}}\,V_{\mathrm{ud/us}}\,\overline{u}_{\nu_\tau}\gamma^\mu(1-\gamma_5)\,u_\tau \mathcal{H}_\mu\,,
\end{equation}
where our lack of knowledge of the precise hadronization mechanism is encoded in the hadronic vector, $\mathcal{H}_\mu$:
\begin{equation} \label{Hmugral}
\mathcal{H}_\mu = \bra \left\lbrace  P(p_i)\right\rbrace_{i=1}^n |\left( \mathcal{V}_\mu - \mathcal{A}_\mu\right)  e^{i\mathcal{L}_{QCD}}|0\ket\,.
\end{equation}
\indent Symmetries help us to decompose $\mathcal{H}_\mu$ depending on the number of final-state pseudoscalar ($P$) mesons, $n$. For three mesons in the final state, this reads:
\begin{eqnarray} \label{Hmu3m}
\mathcal{H}_\mu = V_{1\mu} F_1^A(Q^2,s_1,s_2) + V_{2\mu} F_2^A(Q^2,s_1,s_2) +\nonumber\\
 Q_\mu F_3^A(Q^2,s_1,s_2) + i \,V_{3\mu} F_4^V(Q^2,s_1,s_2)\,,
\end{eqnarray}
and
\begin{eqnarray} \label{VmuQmu}
V_{1\mu} &  = & \left( g_{\mu\nu} - \frac{Q_{\mu}Q_{\nu}}{Q^2}\right) \,
(p_2 - p_1)^{\nu} ,\nonumber\\
V_{2\mu} & = & \left( g_{\mu\nu} - \frac{Q_{\mu}Q_{\nu}}{Q^2}\right) \,
(p_3 - p_1)^{\nu}\,,\nonumber\\
V_{3\mu} & = & \varepsilon_{\mu\nu\varrho\sigma}\,p_1^\nu\, \,p_2^\varrho\, \,p_3^{\sigma} ,\nonumber\\ Q_\mu & = & (p_1\,+\,p_2\,+\,p_3)_\mu \,,\,s_i = (Q-p_i)^2\,.
\end{eqnarray}
\indent $F_i$, $i=1,2,3$, correspond to the axial-vector current ($\mathcal{A}_\m$) while $F_4$ drives the vector current ($\mathcal{V}_\m$). The form factors $F_1$ and $F_2$ have a transverse structure in the total hadron momenta, $Q^\mu$, and drive a $J^P=1^+$ transition. The pseudoscalar form factor, $F_3$, vanishes with the mass of the Goldstone bosons (chiral limit) and, accordingly, gives a tiny contribution. This is as far as we can go without model assumptions, that is, we are still not able to derive the $F_i$ from QCD \cite{Fritzsch:1973pi}. Our claim is that the most adequate way out is the use of a phenomenologically motivated theory that, in the energy region spanned by tau decays, resembles QCD as much as possible. For this, the approximate symmetries of QCD are indeed useful. They rule what is the theory to be used in its very low-energy domain and guide the construction of higher-energy theories.
\section{Resonance Chiral Theory} \label{Theory}
\indent Weinberg's Theorem \cite{Weinberg:1978kz} states that a description in terms of the relevant degrees of freedom for a given process that fulfills all general fundamental QFT-symmetries plus those characteristic of the theory at hand will give the most general S-matrix elements -and thus observables- corresponding to the underlying theory. The principles behind effective-field theories \cite{Georgi:1994qn} would indicate this to be advisable, moreover.\\
\indent $\chi PT$ is the effective theory for very low-energy QCD \cite{ChPT}. However, the tau mass value prevents it to explain its decays through all the energy region they span \cite{Colangelo:1996hs}. One then needs a theory fulfilling the chiral symmetry constraints that includes explicitly the light-flavoured resonances. For it to resemble QCD at larger energies, however, one needs to ensure the right asymptotic behaviour for Green functions and related form factors \cite{BrodskyLepage}. Finally, for it to be useful, one should demand the existence of an expansion parameter to build any perturbative computation upon it. Long ago, it was proposed the inverse of the number of colours of the color gauge group to do this task \cite{Nc}, and has proved to work successfully explaining meson phenomenology \cite{NcToni}.\\
\indent The theory satisfying all these requirements is Resonance Chiral Theory ($R\chi T$) \cite{RChT}. Further studies within it have been presented in this conference in fields as distant as lepton flavor violating $\t$ decays including mesons \cite{Herrero}, and in a lattice evaluation of the hadronic contribution to the anomalous magnetic moment of the muon \cite{lattice}.\\
\indent The relevant part of the R$\chi$T Lagrangian is \cite{RChT}, \cite{vap}, \cite{vvp}, \cite{paper}:
\begin{eqnarray} \label{Full_Lagrangian}
& & \mathcal{L}_{R\chi T}=\frac{F^2}{4}\bra u_\mu u^\mu +\chi_+ \ket+\frac{F_V}{2\sqrt{2}}\bra V_{\mu\nu} f^{\mu\nu}_+ \ket \nonumber\\
& & + \frac{i \,G_V}{\sqrt{2}} \bra V_{\mu\nu} u^\mu u^\nu\ket + \frac{F_A}{2\sqrt{2}}\bra A_{\mu\nu} f^{\mu\nu}_- \ket +\mathcal{L}_{\mathrm{kin}}^V\nonumber\\
& & + \mathcal{L}_{\mathrm{kin}}^A + \sum_{i=1}^{5}\lambda_i\mathcal{O}^i_{VAP} + \sum_{i=1}^7\frac{c_i}{M_V}\mathcal{O}_{VJP}^i \nonumber\\
& & + \sum_{i=1}^4d_i\mathcal{O}_{VVP}^i + \sum_{i=1}^5 \frac{g_i}{M_V} {\mathcal O}^i_{VPPP} \, ,
\end{eqnarray}
where all couplings are real, being $F$ the pion decay constant in the chiral limit. The notation is that of Ref.~\cite{RChT}. $P$ stands for the lightest pseudoscalar mesons and $A$ and $V$ for the (axial)-vector mesons. Furthermore, all couplings in the last two lines are defined to be dimensionless. The subindex of the operators stands for the kind of vertex described, i.e., ${\mathcal O}^i_{VPPP}$ gives a coupling between one Vector and three Pseudoscalars. For the explicit form of the operators in the last line, see \cite{vap}, \cite{vvp}, \cite{paper}. For the matching of $2$ and $3$-point Green Functions in the OPE of QCD and in R$\chi$T and for requiring a Brodsky-Lepage behaviour to the corresponding form factors see \cite{RChT}, \cite{vap}, \cite{vvp}, \cite{amo}, \cite{knecht},  \cite{consistent}, \cite{Rosell:2006dt}, \cite{Portoles:2006nr}, \cite{Mateu:2007tr}, \cite{Pich:2008jm}. Finally, resonance widths are included in a quantum field theory sound way, as explained in \cite{width}. In Ref.~\cite{paper} we give for the first time an axial-vector width according to that definition.
\section{ASYMPTOTIC BEHAVIOUR AND QCD CONSTRAINTS} \label{Asymptotic_behaviour}
\indent There are 24 unknown couplings in $\mathcal{L}_{R\chi T}$ (\ref{Full_Lagrangian}) that may appear in the calculation of three meson decays of the $\t$.  The number of unknowns is reduced when computing the Feynman diagrams involved. There only appear $F_V$, $G_V$, three combinations of the $\lbrace \lambda_i\rbrace_{i=1}^5$, four of the $\lbrace c_i\rbrace_{i=1}^7$, two of the $\lbrace d_i\rbrace_{i=1}^4$ and four of the $\lbrace g_i\rbrace_{i=1}^5$. The number of free parameters has been reduced from 24 to 15.\\
\indent We require the form factors of the $\mathcal{A}^\m$ and $\mathcal{V}^\m$ currents into $KK\p$ modes vanish at infinite transfer of momentum. As a result, we obtain constraints \cite{paper}, \cite{Roig:2007yp} among all axial-vector current couplings but $\lambda_0$, that are also the most general ones satisfying the demanded asymptotic behaviour in $\t\to3\p\n_\t$, studied along these lines in \cite{t3p}. Proceeding analogously with the vector current form factor results in five additional restrictions \cite{paper}. From the 24 initially free couplings in Eq.(\ref{Full_Lagrangian}), only five remain free: $c_4$, $c_1\,+\,c_2\,+\,8\,c_3\,-\,c_5$, $d_1\,+\,8\,d_2\,-\,d_3$, $g_4$ and $g_5$. After fitting $\Gamma(\omega\to3\pi)$ -using some of the relations in Ref. \cite{vvp}- only $c_4$ and $g_4$ remain unknown. We obtained them from BaBar data on $e^+e^-\to KK\pi$.
\section{Phenomenology} \label{Pheno}
\subsection{$\t\to3\p\n_\t$}
\begin{figure}[h] \label{Fig3pi}
  \begin{center}
\includegraphics[scale=0.3]{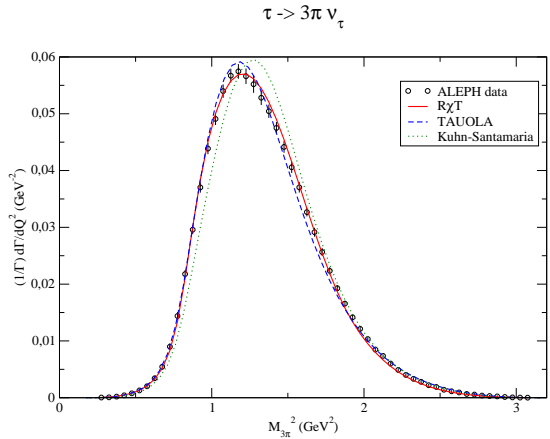}
\end{center}

\caption{\small{A comparison of different parameterizations to the ALEPH normalized spectral function of $\tau\to 3\pi \nu_\tau$ \cite{aleph0}.}}
\label{K+K-pi-}
\end{figure}

\indent In this communication, we concentrate on the semileptonic decays of the $\t$ into $3\p$ and $KK\p$ modes. The $\p\p$ \cite{pipi} and $K\pi$ \cite{Kpi} decay channels have already been analyzed successfully within our framework. For other devoted studies, see \cite{p06} and references therein.\\
\indent We present here an updated analyses for the $3\pi$ mode \cite{paper} that we confront to ALEPH data \cite {aleph0}, to the original work of K\"uhn and Santamar\'ia \cite{KS} and to the parameterization included in TAUOLA \cite{TAUOLA}. In the Figure \ref{Fig3pi} we see that our parameterization -although depending on just one free parameter- is the one working better. 
 While in section \ref{Theory} we saw that our approach is best motivated theoretically, we have checked now that it is appropriately supported by phenomenology, as well.
\subsection{$\t\to KK\p\n_\t$}
\begin{figure}[h] \label{Fig_K+K-pi-}
  \begin{center}
\includegraphics[scale=0.3,angle=-90]{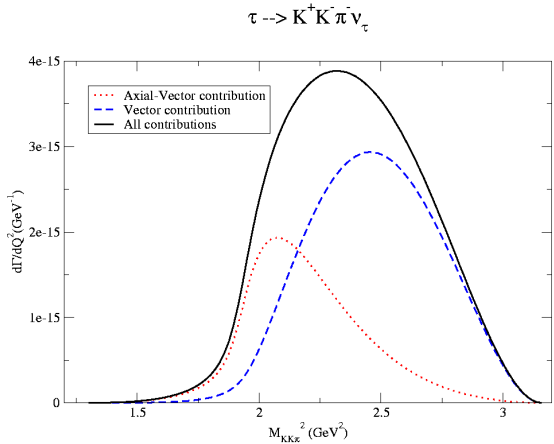}
\end{center}

\caption{\small{Spectral function of $\tau\to K^+K^-\pi^-\nu_\tau$ obtained with the set of parameters (\ref{coups_final}).}}
\label{K+K-pi-}
\end{figure}

\indent In this conference we were extremely glad to see the nice Belle preliminary data on these modes \cite{Yoko}. For the moment, however, we lack of experimental data on the spectral function of $\t\to KK\p\n_\t$ publicly available.\\
\indent Fortunately, it is possible to relate the latter to the $I=1$ component of $e^+e^-\to KK\pi$~\cite{Roig:2008ev}. The issue is discussed in detail in our article \cite{paper}. We conclude that \cite{Roig:2008xt} the measured isovector component in $e^+e^-\to K_S K^{\pm} \pi^{\mp}$ does not provide the total isovector component for the (partially) inclusive decay $e^+e^-\to KK\pi$. Once some assumptions are used, one can employ $CVC$ to relate $e^+e^-\to KK\pi$ to $\t \to KK\pi\nu_\t$ \cite{mal}. We fitted our expressions to BaBar data \cite{babar2} obtaining $c_4=-0.044\pm 0.005$. Our results for the decay widths of the considered channels are consistent with the PDG values \cite{PDG2006}. To confront our predictions for the spectra with the forthcoming experimental data is a necessary task.\\
\indent We have obtained several sets of allowed values for those quantities that were still free after the high-energy conditions were imposed \cite{paper}.
We get two predictions:
\begin{itemize}
\item The ratio of the vector current to all contributions is, for all charge modes:
\begin{equation}
\frac{\Gamma_V\left(\tau\to KK\pi\n_\t\right)}{\Gamma\left(\tau\to KK\pi\n_\t\right)} \sim 0.5.
\end{equation}
\item The spectra of the two independent $KK\pi$ decay modes of the $\tau$. We show in Figure \ref{Fig_K+K-pi-} that one for $\tau\to K^+K^-\pi^-\n_\t$. The shape of the spectral function is similar for $\tau\to K^-K^0\pi^0\n_\t$. In both cases, the axial-vector current dominates the low-energy region, while the vector one peaks at higher values of $Q^2$.
\end{itemize}
The plot presented here corresponds to:
\begin{eqnarray} \label{coups_final}
M_{a_1} = 1.17 \,\,\mathrm{GeV},\,\,c_4 = -0.04,\,\,g_4=-0.5.
\end{eqnarray}

\section{K\"uhn-Santamar\'ia-like models}
\indent In all the presentations of our work, we have devoted special care to differentiate between the original K\"uhn-Santamar\'ia work \cite{KS} and those done in analogy for other decay modes. While \cite{KS} describes very nicely the two- and three-pion modes (just see the amount of work needed to improve it) we can reproached them -from the theory side- three aspects:
\begin{itemize}
\item Although widely believed, Breit-Wigner functions are just the easiest option to formulate resonance exchange and its link with QCD is far from proved.
 \item The proposed off-shell widths are completely ad-hoc and being mass and width of particles definition-dependent objects, it is desirable to follow a QFT-approach to study the axial-vector resonances.
 \item It fails to reproduce the $\mathcal{O}(p^4)\,\chi PT$ result at low energies \cite{p00}, \cite{t3p}, which is numerically much less important than in other channels, being all pions in this case.
\end{itemize}
 These three topics may be criticized in the same way for all subsequent works done along these lines.\\
\indent In the $KK \pi$ channels, there are more (particularly important) aspects to be improved:
\begin{itemize}
 \item The authors of Refs. \cite{fm} needed to include noticeable different masses and widths for the $\rho'$ resonance in the vector and axial-vector currents.
\item The inclusion of three multiplets in the vector form-factors and two in the axial-vector ones is unnatural. One may cut the spectrum and heavier integrated-out resonances effect will be encoded in those couplings of the lighter, still active ones. Proceeding differently in the various form factors has no theoretical reason behind.
 \item Throughout our work \cite{paper}, we have noticed that these studies lack from several contributions that happen to be non-negligible in our formalism, namely, they only account for $\rho^{0,-}$ and $K^{*0}$ being exchanged in one channel in the modes $K^+K^-\pi^-$, $K^-K^0\pi^0$, $K^- \pi^- \pi^+$ and $\overline{K}^0\pi^0 \pi ^-$. That is, they are only including these exchanges either in $V_1^\mu$ or $V_2^\mu$ (\ref{VmuQmu}), while one naturally obtains them in both.
\end{itemize}
\indent Moreover, the CLEO collaboration \cite{cleo} reported that it was not possible to fit their data using the parameterizations in Refs. \cite{fm}. They modified it by including two additional parameters in order to obtain a good fit, but at the prize of violating a property of the strong interactions, namely the Wess-Zumino \cite{WZW} normalization emanating from the chiral anomaly of QCD, as was put forward in Ref. \cite{p04}.

\section{Conclusions and Outlook}
In his talk, Z.~Was \cite{Was} stated that ''Hadronic currents need to remain experiments' property, in case experiments wish so''. We have seen with delight throughout this conference, that experimentalists do wish, as they presented many interesting measurements and announced forthcoming, complementary ones.\\
\indent Let me point out, notwithstanding, that we will only get full benefit from their work if we are able to relate their measurements to the theory behind, that -electroweak effects factorized- is QCD.\\
\indent We have highlighted in section \ref{Theory} the guidelines of our theory, that includes every relevant (and known) piece of QCD: We preserve the correct chiral limit at low energies that gives the right normalization to our form factors, we employ large-$N_C$ QCD arguments to use our theory in terms of mesons and it fulfills the high-energy conditions of the fundamental theory at the mesonic level. While we remain predictive by including only the lightest multiplet of axial-vector and vector resonances, our Lagrangian can easily be extended in a systematic way to account for the contribution of higher resonance states, whose parameters may be fitted to experiment.\\
\indent In section \ref{Pheno} we have explicitly showed that there are not only theoretical reasons to prefer our parameterization. It works better for the $3\p$ and $KK\p$ modes. Immediately, after, we have reviewed all aspects of the KS-models that need improvement. Noteworthy, all of them get corrected within our framework.\\
\indent Therefore, we strongly suggest the implementation of our form factors in the TAUOLA library \cite{TAUOLA}. For those modes we do not have calculated yet, the expressions in TAUOLA are the best guidance at the moment.\\
\indent We have already worked out the $\t\to K\p\p\n_\t$ decay, for which preliminary data was shown in this conference \cite{Lee}. This way, we will be able to help the simultaneous extraction of $V_{us}$ and $m_s$ using tau decays \cite{Gamiz:2004ar}. We plan to tackle all three meson decays of the tau along the lines described here.\\
\indent Our Lagrangian approach can also be used to compute exclusively $e^+e^-$ into hadrons. This could be useful for improving the hadronic matrix elements in PHOKHARA \cite{PHOKHARA}.\\
\indent Finally, we want to stress how important is that the MonteCarlos collect all this QCD-features, as they constitute the link between theory and experiment. In the Working Group on Radiative Corrections and Generators for Low Energy Hadronic Cross Section and Luminosity held in Frascati last April, there was unanimity in that understanding the hadronic contribution both in $e^+e^-$ and in $\t$ decays is mandatory for the MonteCarlo's employed at the $\t$-charm- and $B$-factories. This task will be of great help also in LHC and any future collider.


\begin{thebibliography}{9}
\bibitem{Pich:2007cu}
  A.~Pich,
  Nucl.\ Phys.\ Proc.\ Suppl.\  {\bf 169} (2007) 393
  [arXiv:hep-ph/0702074].
  arXiv:0806.2793 [hep-ph].

\bibitem{Fritzsch:1973pi}
  H.~Fritzsch, M.~Gell-Mann and H.~Leutwyler,
  Phys.\ Lett.\  B {\bf 47} (1973) 365.
%
\bibitem{Weinberg:1978kz}
  S.~Weinberg,
  Physica A {\bf 96} (1979) 327.

\bibitem{Georgi:1994qn}
  H.~Georgi,
  Ann.\ Rev.\ Nucl.\ Part.\ Sci.\  {\bf 43} (1993) 209.
%
  A.~V.~Manohar,
  arXiv:hep-ph/9606222.
 Published in 'Schladming 1996, Perturbative and nonperturbative aspects of quantum field theory' 311-362. 
%
  A.~Pich,
  arXiv:hep-ph/9806303. Published in 'Les Houches 1997, Probing the standard model of particle interactions', Pt. 2 949-1049.

\bibitem{ChPT}
J.~Gasser and H.~Leutwyler,
  Annals Phys.\  {\bf 158} (1984) 142.
  Nucl.\ Phys.\  B {\bf 250} (1985) 465.

\bibitem{Colangelo:1996hs}
  G.~Colangelo, M.~Finkemeier and R.~Urech,
  Phys.\ Rev.\  D {\bf 54} (1996) 4403
  [arXiv:hep-ph/9604279].

\bibitem{BrodskyLepage}
S.~J.~Brodsky and G.~R.~Farrar,
  Phys.\ Rev.\ Lett.\  {\bf 31} (1973) 1153.
 G.~P.~Lepage and S.~J.~Brodsky,
  Phys.\ Rev.\  D {\bf 22} (1980) 2157.

\bibitem{Nc}
G.~'t Hooft,
  Nucl.\ Phys.\  B {\bf 72} (1974) 461.
  Nucl.\ Phys.\  B {\bf 75} (1974) 461.
E.~Witten,
  Nucl.\ Phys.\  B {\bf 160} (1979) 57.

\bibitem{NcToni}
A.~V.~Manohar,
  arXiv:hep-ph/9802419. Published in 'Les Houches 1997, Probing the standard model of particle interactions, Pt. 2'.
A.~Pich,
  arXiv:hep-ph/0205030. Published in 'Tempe 2002, Phenomenology of large N(c) QCD' 239-258.

\bibitem{RChT}
 G.~Ecker, J.~Gasser, A.~Pich and E.~de Rafael,
  Nucl.\ Phys.\  B {\bf 321} (1989) 311.
G.~Ecker, J.~Gasser, H.~Leutwyler, A.~Pich and E.~de Rafael,
  Phys.\ Lett.\  B {\bf 223} (1989) 425.

\bibitem{Herrero}
E.~Arganda, M.~J.~Herrero and J.~Portol\'es,
  JHEP {\bf 0806} (2008) 079
  [arXiv:0803.2039 [hep-ph]].
M.~J.~Herrero, these proceedings.

\bibitem{lattice}
C.~Aubin and T.~Blum,
  PoS {\bf LAT2005} (2006) 089
  [arXiv:hep-lat/0509064].
T.~Blum, these proceedings.

\bibitem{vap}
 V.~Cirigliano, G.~Ecker, M.~Eidem\"uller, A.~Pich and J.~Portol\'es,
  Phys.\ Lett.\  B {\bf 596} (2004) 96
  [arXiv:hep-ph/0404004].

\bibitem{vvp}
 P.~D.~Ruiz-Femen\'ia, A.~Pich and J.~Portol\'es,
  JHEP {\bf 0307} (2003) 003
  [arXiv:hep-ph/0306157].

\bibitem{paper}
D.~G\'omez-Dumm, P.~Roig, A.~Pich, J.~Portol\'es, to appear.

\bibitem{amo}
 G.~Amor\'os, S.~Noguera and J.~Portol\'es,
  Eur.\ Phys.\ J.\  C {\bf 27} (2003) 243
  [arXiv:hep-ph/0109169].

\bibitem{knecht}
M.~Knecht and A.~Nyffeler,
  Eur.\ Phys.\ J.\  C {\bf 21} (2001) 659
  [arXiv:hep-ph/0106034].

\bibitem{consistent}
V.~Cirigliano, G.~Ecker, M.~Eidem\"uller, R.~Kaiser, A.~Pich and J.~Portol\'es,
  Nucl.\ Phys.\  B {\bf 753} (2006) 139
  [arXiv:hep-ph/0603205].

\bibitem{Rosell:2006dt}
  I.~Rosell, J.~J.~Sanz-Cillero and A.~Pich,
  JHEP {\bf 0701} (2007) 039
  [arXiv:hep-ph/0610290].

\bibitem{Portoles:2006nr}
  J.~Portol\'es, I.~Rosell and P.~Ruiz-Femen\'ia,
  Phys.\ Rev.\  D {\bf 75} (2007) 114011
  [arXiv:hep-ph/0611375].

\bibitem{Mateu:2007tr}
  V.~Mateu and J.~Portol\'es,
  Eur.\ Phys.\ J.\  C {\bf 52} (2007) 325
  [arXiv:0706.1039 [hep-ph]].

\bibitem{Pich:2008jm}
 A.~Pich, I.~Rosell and J.~J.~Sanz-Cillero,
  JHEP {\bf 0807} (2008) 014
  [arXiv:0803.1567 [hep-ph]].

\bibitem{width}
D.~G\'omez Dumm, A.~Pich and J.~Portol\'es,
  Phys.\ Rev.\  D {\bf 62} (2000) 054014
  [arXiv:hep-ph/0003320].

\bibitem{Roig:2007yp}
 P.~Roig, AIP Conf.\ Proc.\  {\bf 964} (2007) 40
  [arXiv:0709.3734 [hep-ph]].

\bibitem{t3p}
D.~G\'omez Dumm, A.~Pich and J.~Portol\'es,
  Phys.\ Rev.\  D {\bf 69} (2004) 073002
  [arXiv:hep-ph/0312183].

\bibitem{pipi}
F.~Guerrero and A.~Pich,
  Phys.\ Lett.\  B {\bf 412} (1997) 382
  [arXiv:hep-ph/9707347].
A.~Pich and J.~Portol\'es,
  Phys.\ Rev.\  D {\bf 63} (2001) 093005
  [arXiv:hep-ph/0101194].
  Nucl.\ Phys.\ Proc.\ Suppl.\  {\bf 121} (2003) 179
  [arXiv:hep-ph/0209224].

\bibitem{Kpi}
M.~Jamin, A.~Pich and J.~Portol\'es,
  Phys.\ Lett.\  B {\bf 640} (2006) 176
  [arXiv:hep-ph/0605096].
  Phys.\ Lett.\  B {\bf 664} (2008) 78
  [arXiv:0803.1786 [hep-ph]].

\bibitem{p06}
J.~Portol\'es,
  Nucl.\ Phys.\ Proc.\ Suppl.\  {\bf 169} (2007) 3
  [arXiv:hep-ph/0702132].

\bibitem{aleph0}
R.~Barate {\it et al.}  [ALEPH Collaboration],
  Eur.\ Phys.\ J.\  C {\bf 4} (1998) 409.

\bibitem{KS}
J.~H.~K\"uhn and A.~Santamar\'\i{}a,
Z.\ Phys.\ C {\bf 48} (1990) 445.

\bibitem{TAUOLA}
S.~Jadach, Z.~Was, R.~Decker and J.~H.~K\"uhn, Comput.\ Phys.\ Commun.\  {\bf 76} (1993) 361.

\bibitem{Yoko}
Y.~Usuki, Talk ''$\tau \to K^* K \nu_\tau$ at Belle'', these proceedings.


\bibitem{Roig:2008ev}
P.~Roig, 
  arXiv:0810.1255 [hep-ph].
 Nucl.Phys.B, Proc.Suppl.181-182 2008:319-323, 2008.

\bibitem{Roig:2008xt}
  P.~Roig,
  arXiv:0810.2187 [hep-ph]. To appear in the Proceedings of QCD 08.

\bibitem{mal}M.~Davier, S.~Descotes-Genon, A.~H\"ocker, B.~Malaescu, Z.~Zhang,
Eur.\ Phys.\ J.\  C {\bf 56} (2008) 305 [arXiv:0803.0979].
M.~Davier, these proceedings.

\bibitem{babar2}
B.~Aubert {\it et al.}  [BABAR Collaboration], Phys.\ Rev.\  D {\bf 77} (2008) 092002.

\bibitem{PDG2006}
W.-M. Yao et al., Journal of Physics G \textbf{33}, 1 (2006).

\bibitem{p00}
 J.~Portol\'es,
  Nucl.\ Phys.\ Proc.\ Suppl.\  {\bf 98} (2001) 210
  [arXiv:hep-ph/0011303].

\bibitem{fm}
 M.~Finkemeier and E.~Mirkes,
  Z.\ Phys.\  C {\bf 69} (1996) 243
  [arXiv:hep-ph/9503474].
%
 J.~H.~K\"uhn, E.~Mirkes and M.~Finkemeier,
  [arXiv:hep-ph/9511268].

\bibitem{cleo}F.~Liu  [CLEO Collaboration],
  F.~Liu  [CLEO Collaboration],
  eConf {\bf C0209101} (2002) TU07
  [Nucl.\ Phys.\ Proc.\ Suppl.\  {\bf 123} (2003) 66]
  [arXiv:hep-ex/0209025].
T.~E.~Coan {\it et al.}  [CLEO Collaboration],
  Phys.\ Rev.\ Lett.\  {\bf 92} (2004) 232001
  [arXiv:hep-ex/0401005].

\bibitem{WZW}
 J.~Wess and B.~Zumino,
  Phys.\ Lett.\  B {\bf 37} (1971) 95.
E.~Witten,
  Nucl.\ Phys.\  B {\bf 223} (1983) 422.

\bibitem{p04}
 J.~Portol\'es,
  Nucl.\ Phys.\ Proc.\ Suppl.\  {\bf 144} (2005) 3
  [arXiv:hep-ph/0411333].

\bibitem{Was}
Z.~Was, talk ''Simulation with TAUOLA and PHOTOS'', these proceedings.

\bibitem{Lee}
M.~J.~Lee, 
 these proceedings.

\bibitem{Gamiz:2004ar}
  E.~G\'amiz, M.~Jamin, A.~Pich, J.~Prades and F.~Schwab,
  Phys.\ Rev.\ Lett.\  {\bf 94} (2005) 011803
  [arXiv:hep-ph/0408044].

\bibitem{PHOKHARA}
  G.~Rodrigo, H.~Czyz, J.~H.~K\"uhn and M.~Szopa,
  Eur.\ Phys.\ J.\  C {\bf 24} (2002) 71
  [arXiv:hep-ph/0112184].
\end{thebibliography}
\end{document}